\documentclass{ifacmtg}
\usepackage[dcucite]{harvard}

\usepackage{amsmath,amssymb,euscript,amscd,graphicx, color, amsfonts, fleqn }

\usepackage{graphicx}

\newtheorem{corollary}{Corollary}[section]
\newtheorem{lemma}{Lemma}[section]

\newtheorem{definition}{Definition}[section]

\def\bbr{{\mathbb R}}
\definecolor{mybackgroundcolor}{rgb}{1,0,1}
\definecolor{plum}{rgb}{.5,0,.5}
\definecolor{dkred}{rgb}{.5,0,0}
\definecolor{ddkred}{rgb}{.35,0,0}
\definecolor{dkblue}{rgb}{.1,0,.6}
\definecolor{ddkblue}{rgb}{0,0,.25}
\definecolor{dkgreen}{rgb}{0,.5,0}
\definecolor{ddkgreen}{rgb}{0,.35,0}
\definecolor{dkgr2}{rgb}{0,.57,0}
\definecolor{dkgr3}{rgb}{0,.64,0}
\definecolor{dkgr4}{rgb}{0,.71,0}

\pagecolor{white}

\newcommand{\be}{\begin{equation}}
\newcommand{\ee}{\end{equation}}
\newcommand{\bd}{\begin{displaymath}}
\newcommand{\ed}{\end{displaymath}}
\newcommand{\bea}{\begin{eqnarray}}
\newcommand{\eea}{\end{eqnarray}}
\newcommand{\rhp}{\mathbb{C}_{+}}
\newcommand{\Zp}{\mathbb{Z}_{+}}

\newcommand{\Hi} {{\cal H}^\infty}

\begin{document}
\runauthor{Cicero, Caesar and Vergil}
\begin{frontmatter}
\title{Remarks on $\Hi$ Controller Design for SISO Plants
with Time Delays\thanksref{MIRG}}
\author[Gumussoy]{Suat G\"um\"u\c{s}soy}
\author[Ozbay]{Hitay \"Ozbay}
\address[Gumussoy]{was
with Dept. of Electrical and Computer Eng.,\\ Ohio State
University, Columbus, OH 43210, U.S.A.; current affiliation: MIKES
Inc., Akyurt, Ankara TR-06750, Turkey, {\tt
suat.gumussoy@mikes.com.tr}}
\address[Ozbay]{Dept. of
Electrical and Electronics Eng.,\\ Bilkent University, Bilkent,
Ankara TR-06800, Turkey, \\on leave from \\Dept. of Electrical and
Computer Eng., \\ Ohio State University, Columbus, OH 43210,
U.S.A., {\tt hitay@bilkent.edu.tr, ozbay@ece.osu.edu}}
\thanks[MIRG]{This work was supported in part by
the European Commission (contract no. MIRG-CT-2004-006666) and by
T\"UB\.{I}TAK (grant no. EEEAG-105E065).}

\begin{abstract}
The skew Toeplitz approach is one of the well developed methods to
design $\Hi$ controllers for infinite dimensional systems. In
order to be able to use this method the plant needs to be
factorized in some special manner. This paper investigates the
largest class of SISO time delay systems for which the special
factorizations required by the skew Toeplitz approach can be done.
Reliable implementation of the optimal controller is also
discussed. It is shown that the finite impulse response (FIR)
block structure appears in these controllers not only for plants
with I/O delays, but also for general time-delay
plants.
\end{abstract}
\end{frontmatter}

\section{Introduction}
There are many well-developed techniques for finding $\Hi$ optimal
and suboptimal controllers for systems with time delays. In
particular, when the plant is a dead-time system: $ e^{-hs}P_0(s)$
where $P_0$ is a rational SISO plant, the optimal $\Hi$ control
problem is solved by \cite{ZK87}, \cite{FTZ86}, using operator
theoretic methods; see also \cite{S89}, \cite{O90} and their
references. State-space solution to the same problem is given in
\cite{T97}, and \cite{MZ00}. Notably, \cite{MZ00} used
$J$-spectral factorization approach to solve the MIMO version of
the problem. Moreover they showed the finite impulse response
(FIR) structure appearing in the reliable implementation of the
$\Hi$ controllers for dead-time systems. \cite{MM04} extended this
result to the multi-delay dead-time systems (input/output delay
case).

A closed-form controller formula is obtained  by \cite{KY03} for
the sensitivity minimization problem involving pseudorational
plants. For more general infinite dimensional plants a solution is
given by \cite{FOT}. Their approach needs inner-outer
factorization of the plant. \cite{TO95} simplified this method and
brought into a compact form.

 \cite{KPhD05} obtained an expression for the optimal $\Hi$ controller  for
the plants that can be expressed as a cascade connection of a
finite-dimensional generalized plant and a scalar inner function.
As it was done by \cite{M03}, the solution is reduced to solving
two algebraic Riccati equations and an additional one-block
problem. Moreover, \cite{KPhD05} gave the inner-outer
factorizations of stable pseudorational systems.

In our study, we determine the largest class of time-delay systems
(TDS) for which the Skew-Toeplitz approach of \cite{FOT} is
applicable. In order to use this method it is necessary to do
inner-outer factorizations of the plant. An additional assumption
is that the infinite dimensional plant has finitely many unstable
zeros or poles. In this paper, we give necessary and sufficient
conditions for TDS to have finitely many unstable zeros or poles.
We classify the TDS and give conditions such that the desired
factorization is possible. For admissible plants, the
factorization is given and optimal $\Hi$ controller is obtained.
The unstable pole-zero cancellation in the optimal controller
expression of \cite{TO95} is eliminated. This way we establish the
link between \cite{TO95} and \cite{MZ00} by showing the FIR
structure appearing in $\Hi$ controllers for not only dead-time
plants, but also for more general TDS.

\vspace*{-0.35cm}

\section{Preliminary Definitions and Results} \label{section:prework}

In \cite{FOT,TO95}, it is assumed that the plant is in the form
\be
\label{eq:plantP1}
\hat{P}(s)=\frac{\hat{m}_n(s)\hat{N}_o(s)}{\hat{m}_d(s)}
\ee
where $\hat{m}_n(s)$ is inner, infinite dimensional and
$\hat{m}_d(s)$ is inner, finite dimensional and $\hat{N}_o(s)$ is
outer, possibly infinite dimensional. The optimal $\Hi$
controller, $\hat{C}_{opt}$, stabilizes the feedback system and
achieves the minimum $\Hi$ cost, $\hat{\gamma}_{opt}$:
\be
\label{eq:wsm1} \hat{\gamma}_{opt}=\left\| \left[\begin{array}{c}
  \hat{W}_1(1+\hat{P}\hat{C}_{opt})^{-1} \\
  \hat{W}_2\hat{P}\hat{C}_{opt}(1+\hat{P}\hat{C}_{opt})^{-1}
\end{array} \right]\right\|_\infty
\ee
where $\hat{W}_1$ and $\hat{W}_2$ are finite dimensional weights
of the mixed sensitivity minimization problem.

Recently, the optimal $\Hi$ control problem is solved by
\cite{GO04} for systems with infinitely many unstable poles and
finitely many unstable zeros by using the duality with the problem
(\ref{eq:wsm1}). In this case, the plant has a factorization
\be
\label{eq:plantP2}
\tilde{P}(s)=\frac{\tilde{m}_d(s)\tilde{N}_o(s)}{\tilde{m}_n(s)}
\ee
where $\tilde{m}_n$ is inner, infinite dimensional,
$\tilde{m}_d(s)$ is finite dimensional, inner, and
$\tilde{N}_o(s)$ is outer, possibly infinite dimensional. For this
dual problem, the optimal controller, $\tilde{C}_{opt}$, and
minimum $\Hi$ cost, $\tilde{\gamma}_{opt}$, are found for the
mixed sensitivity minimization problem
\be \label{eq:wsm2} \tilde{\gamma}_{opt}=\left\|
\left[\begin{array}{c}
  \tilde{W}_1(1+\tilde{P}\tilde{C}_{opt})^{-1} \\
  \tilde{W}_2\tilde{P}\tilde{C}_{opt}(1+\tilde{P}\tilde{C}_{opt})^{-1}
\end{array} \right]\right\|_\infty.
\ee

In this paper we consider general delay systems:
\be \label{eq:preplant}
P(s)=\frac{r_p(s)}{t_p(s)}=\frac{\sum_{i=1}^{n}r_{p,i}(s)e^{-h_i
s}}{\sum_{j=1}^{m}t_{p,j}(s)e^{-\tau_j s}}
\ee
satisfying the assumptions
\begin{enumerate}
    \item[A.1]
    \begin{enumerate}
    \item $r_{p,i}(s)$ and $t_{p,j}(s)$ are polynomials with
    real coefficients;
    \item $h_i$, $\tau_j$ are rational numbers such
    that $0\leq h_1<h_2<\ldots<h_n$, and $0\leq
    \tau_1<\tau_2<\ldots<\tau_m$, with $h_1\geq\tau_1$;
    \item define the polynomials $r_{p,i_{max}}$ and $t_{p,j_{max}}$ with largest
    polynomial degree in $r_{p,i}$ and $t_{p,j}$ respectively
    (the smallest index if there is more than one),
    then, ${\textrm{deg}\{r_{p,i_{max}}(s)\}}
    \leq {\textrm{deg}\{t_{p,{j_{max}}}(s)\}}$
    and
    $h_{i_{max}}\geq\tau_{j_{max}}$ where $\textrm{deg}\{.\}$
    denotes the degree of the polynomial;
    \end{enumerate}
    \item[A.2] $P$ has no imaginary axis zeros or poles;
    \item[A.3] $P$ has finitely  many unstable poles or zeros, or
    equivalently $r_{p}(s)$ or $t_{p}(s)$ has finitely many zeros in
    $\rhp$;
    \item[A.4] $P$ can be written in the form of
    (\ref{eq:plantP1}) or (\ref{eq:plantP2}).
\end{enumerate}

Conditions stated in $A.1$ are not restrictive. In most cases
$A.2$ can be removed if the weights are chosen in a special
manner. The conditions $A.3-A.4$ come from the Skew-Toeplitz
approach.
 It is not easy to check
assumptions $A.3-A.4$, unless a quasi-polynomial root finding
algorithm is used.  We will give a necessary and sufficient
condition to check the assumption $A.3$ in section
\ref{section:finiteassumption} and give conditions to check the
assumption $A.4$  in section \ref{section:plantfactorization}.

By simple rearrangement, $P$ can be written as,
\be
\label{eq:seqplant} P(s)=\frac{R(s)}{T(s)}=\frac{\sum_{i=1}^n
R_i(s)e^{-h_i s}}{\sum_{j=1}^m T_j(s)e^{-\tau_j s}}
\ee
where $R_i$ and $T_j$ are finite dimensional, stable, proper
transfer functions. The assumptions $A.1-A.4$ and rearrangement of
the plant are illustrated on the following example. Consider the
system
\bea
\label{example:IFplant}
\nonumber \dot{x}_1(t)&=&-x_1(t-0.2)-x_2(t)+u(t)+2u(t-0.4),\\
\nonumber \dot{x}_2(t)&=&5x_1(t-0.5)-3u(t)+2u(t-0.4),\\
y(t)&=&x_1(t).
\eea
whose transfer function is in the form
\bea \label{example:IFtf}
\nonumber P(s)&=&\frac{r_{p}(s)}{t_{p}(s)}=\frac{\sum_{i=1}^2
r_{p,i}(s)e^{-h_is}}
{\sum_{i=1}^3 t_{p,i}(s)e^{-\tau_is}}, \\
&=&\frac{(s+3)e^{-0s}+2(s-1)e^{-0.4s}}{s^2e^{-0s}+se^{-0.2s}+5e^{-0.5s}}.
\eea
Note that $r_{p,i}$ and $t_{p,j}$ are polynomials with real
coefficients, delays are nonnegative with increasing order. By
$i_{max}=1$ and $j_{max}=1$, $h_1=0\geq\tau_1=0$ and
${\textrm{deg}\{r_{p,1}(s)\}}=1\leq
{\textrm{deg}\{t_{p,1}(s)\}}=2$. Therefore, assumption $A.1$ is
satisfied. The plant, $P$ has no imaginary axis poles or zeros
(assumption $A.2$). The denominator of the plant, $t_p(s)$ has
finitely many unstable zeros at $0.4672\pm1.8890j$, whereas
$r_p(s)$ has infinitely many unstable zeros converging to
$1.7329~-~(5k+2.5)\pi j$ as $k\rightarrow\infty$. Therefore, plant
has finitely many unstable poles satisfying assumption $A.3$. One
can show that the plant can be factorized as (\ref{eq:plantP1}).
 In this example we have
\be
\label{example:IFRT} P=\frac{R}{T}=\frac{\sum_{i=1}^2
R_i(s)e^{-h_is}}{\sum_{i=1}^3 T_j(s)e^{-\tau_is}}
\ee
where
\bea
 \nonumber R_i(s)&=&\frac{r_{p,i}(s)}{(s+1)^2},
 \quad\textrm{and}\quad T_j=\frac{t_{p,j}(s)}{(s+1)^2}
\eea
are stable proper finite dimensional transfer functions. Below we
give conditions such that $A.3-A.4$ can be checked easily.


\vspace*{-0.5cm}
\subsection{Time Delay Systems with Finitely Many Unstable
Zeros or Poles} \label{section:finiteassumption} \vspace*{-0.25cm}
\begin{definition} \label{def:TDS}
Consider $R(s)=\sum_{i=1}^n R_i(s)e^{-h_i s}$  where each $R_i$ is
a rational, proper, stable transfer function with real
coefficient, and $0\leq h_1<h_2<\ldots<h_n$. Let relative degree
of $R_i(s)$ be $d_i$, then
\begin{itemize}
\item[(i)] if $d_1<\max{\{d_2,\ldots,d_n\}}$, then $R(s)$ is a
retarded-type time-delay system (RTDS),
\item[(ii)] if $d_1=\max{\{d_2,\ldots,d_n\}}$, then $R(s)$ is a
neutral-type time-delay system (NTDS),
\item[(iii)] if $d_1>\max{\{d_2,\ldots,d_n\}}$, then $R(s)$ is an
advanced-type time-delay system (ATDS).
\end{itemize}
\end{definition}
\vspace*{-0.2cm} Note that if $R$ and $T$ are ATDS, plant has
always infinitely many unstable zeros and poles which is not a
valid plant for Skew-Toeplitz approach. It is well-known that RTDS
has finitely unstable zeros on the right-half plane, \cite{BC63}.
Therefore, we will give a necessary and sufficient condition to
check whether a NTDS has finitely many or infinitely many unstable
zeros with the following lemma: \vspace*{-0.2cm}\begin{lemma}
\label{lemma:finitezeros} Assume that $R(s)$ is a NTDS with no
imaginary axis zeros and poles, then the system, $R$, has finitely
many unstable zeros if and only if all the  roots of the
polynomial, $\varphi(r)=1+\sum_{i=2}^n \xi_i
r^{\tilde{h}_i-\tilde{h}_1}$
 has magnitude greater than $1$ where
\bea
\nonumber  \xi_i
 &=&\lim_{\omega\rightarrow\infty}R_i(j\omega)R_1^{-1}(j\omega)
 \quad \forall \;i=2,\dots,n, \\
\nonumber h_i&=&\frac{\tilde{h}_i}{N}, \quad N, \tilde{h}_i
\in\Zp, \; \forall \;i=1,\dots,n.
\eea
{\em Proof}. Since delays are rational numbers, there exist
positive integers $N$ and $\tilde{h}_i$. If $R$ is a NTDS, there
is no root with real part extending to infinity, i.e., \bd
\left|\frac{R(s)e^{h_1s}}{R_1(s)}
\right|_{\sigma\rightarrow\infty}\geq
1-\lim_{\sigma\rightarrow\infty}\sum_{i=2}^n
|\xi_i|e^{-(h_i-h_1)\sigma}>0 \ed where $s=\sigma+j\omega$.
Therefore, NTDS may have infinitely many unstable zeros extending
to infinity in imaginary part with bounded positive real part, see
\cite{BC63}. $R$ has finitely many unstable zeros if and only if
$R(\sigma+j\omega)$ has finitely many zeros as
$\omega\rightarrow\infty$ and $0<\sigma<\sigma_o<\infty$.
Equivalently, $R(\sigma+j\omega)$ has finitely many unstable zeros
if and only if \be \label{eq:Fpoly} \left.
\lim_{\omega\rightarrow\infty}\frac{R(s)}{R_1(s)e^{-h_1s}}\right|
_{s=\sigma_o+j\omega}=1+ \sum_{i=2}^n \xi_i
r^{\tilde{h}_i-\tilde{h}_1}\ee has finitely many unstable zeros
where $r=e^{-\left(\frac{\sigma+j\omega}{N}\right)}$. Let $r_0$ is
the root of (\ref{eq:Fpoly}). Then, \bd |r_o|=e^{-\sigma/N},\quad
\sigma=-N\ln{|r_o|}. \ed Therefore, the system $R$ has finitely
many unstable zeros if and only if all the roots of the polynomial
(\ref{eq:Fpoly}) has magnitude greater than one. Note that if
there exists a root $r_o$ of (\ref{eq:Fpoly}) with $|r_o|\leq1$,
then there are infinitely many unstable zeros of $R$ converging to
$r_{o,k}=-\frac{\ln{|r_o|}}{N}-jN(\angle{r_o}+2\pi k)$ as
$k\rightarrow\infty$ where $k\in\mathbb{Z}$ and $\angle{r_o}$ is
the phase of the complex number $r_o$. \hfill $\Box$
\end{lemma}
\vspace*{-0.2cm}\begin{corollary} \label{cor:Fsystem} The
time-delay system $R$ has finitely many unstable zeros if and only
if  $R$ is a RTDS or $R$ is a NTDS satisfying Lemma
\ref{lemma:finitezeros}.
\end{corollary}
A time delay system with finitely many unstable zeros will be
called an {\em $F$-system}. We define the conjugate of
$R(s)=\sum_{i=1}^n R_i(s)e^{-h_i s}$ as
$\bar{R}(s):=e^{-h_ns}R(-s)M_C(s)$ where $M_C$ is inner, finite
dimensional whose poles are poles of $R$. For the above example,
we have \bd R(s)=\frac{s+3+2(s-1)e^{-0.4s}}{(s+1)^2}\ed where
$h_1=0$, $h_2=0.4$ and $M_C(s)=\left(\frac{s-1}{s+1}\right)^2$.
So, the conjugate of $R(s)$ can be written as, \be
\label{eq:conjR} \bar{R}(s)=\frac{2(s+1)+(s-3)e^{-0.4s}}{(s+1)^2}.
\ee
\begin{corollary} \label{cor:Isystem}
The time-delay system $\bar{R}$ has finitely many unstable zeros
if and only if $R$ is a ATDS or  $R$ is a NTDS with $\bar{R}$
satisfying Lemma \ref{lemma:finitezeros}.
\end{corollary}
The system $R$ whose conjugate $\bar{R}$ has finitely many
unstable zeros is an {\em $I$-system}. Using Corollary
\ref{cor:Fsystem}, an equivalent condition for assumption $A.3$ is
the following.
\begin{corollary} \label{assumption:A3}
Plant (\ref{eq:seqplant}) has finitely many unstable zeros or
poles if and only if $R$ or $T$ is an $F$-system.
\end{corollary}
Using Corollary \ref{assumption:A3}, it is easy to check whether
the plant has finitely many unstable or zeros. After putting the
plant in the form (\ref{eq:seqplant}), if $R$ or $T$ is RTDS, then
assumption $A.3$ is satisfied; if $R$ or $T$ is NTDS and Lemma
\ref{lemma:finitezeros} is satisfied at least for one of them,
then assumption $A.3$ holds.

It is well known that, since $R\in \Hi$, functions in the form $R$
admit inner outer factorizations
\be
\label{eq:mnNo} R=m_nN_o
\ee where $m_n$ is inner  and $N_o$ is
outer. To illustrate this first assume that $R$ is an $F$-system.
By Corollary \ref{cor:Fsystem}, it has finitely many unstable
zeros. Define an inner function $M_R$ whose zeros are unstable
zeros of $R$. Note that $M_R$ is finite dimensional, rational
function. Then, $R$ can be factorized as in (\ref{eq:mnNo}) where
$m_n=M_R$ and $N_o=\frac{R}{M_R}$. Note that unstable zeros of $R$
are cancelled by zeros of $M_R$, therefore $N_o$ is outer and
$m_n$ is inner by construction of $M_R$. Similarly, if $R$ is an
$I$-system, By Corollary \ref{cor:Isystem}, $\bar{R}$ has finitely
many unstable zeros. Define an inner function $M_{\bar{R}}$ whose
zeros are unstable zeros of $\bar{R}$. Using this result, $R$ can
be factorized as in (\ref{eq:mnNo}) where
$m_n=\frac{R}{\bar{R}}M_{\bar{R}}$ and
$N_o=\frac{\bar{R}}{M_{\bar{R}}}$.


\begin{corollary} \label{cor:A3A4}
The plant $P=\frac{R}{T}$ satisfies  $A.3-A.4$ if one of the
following conditions are valid:
\begin{enumerate}
    \item[i)] $R$ is $I$-system and $T$ is $F$-system (IF
    plant),
    \item[ii)] $R$ is $F$-system and $T$ is $I$-system (FI
    plant),
    \item[iii)] $R$ is $F$-system and $T$ is $F$-system (FF
    plant).
\end{enumerate}
\end{corollary}
{\em Proof}. The TDS (\ref{eq:seqplant}) should have finitely many
unstable zeros or poles to apply Skew-Toeplitz approach. By
Corollary \ref{cor:Fsystem}, $R$ or $T$ should be a $F$-system
which covers all the cases except $R$ and $T$ are $I$-systems.
Recall that $P$ (\ref{eq:seqplant}) can be factorized as
\bd
P=\frac{R}{T}=\frac{m_{n,R}N_{o,p}}{m_{n,T}N_{o,T}}=\frac{m_{n,R}}{m_{n,T}}N_o
\ed where $N_o=\frac{N_{o,R}}{N_{o,T}}$ is outer function. Note
that when $R$ is $F$ or $I$-system, $m_{n,R}$ is finite or
infinite dimensional respectively. Similarly, when  $T$ is $F$ or
$I$-system, $m_{n,T}$ is finite or infinite dimensional
respectively. Therefore, the plant (\ref{eq:seqplant}) can be
factorized as (\ref{eq:plantP1}) or (\ref{eq:plantP2}). \hfill
$\Box$

\noindent \textbf{Remarks:}
\begin{enumerate}
\item By Corollary \ref{cor:A3A4}, it is easy to check whether
assumptions $A.3-A.4$ are satisfied or not. For the plant
(\ref{example:IFRT}) in the example, $T$ is a RTDS, by Corollary
\ref{cor:Fsystem}, $T$ is an $F$-system. $R$ is a NTDS and
$\bar{R}$ satisfies Corollary \ref{cor:Isystem}, therefore, $R$ is
a $I$-system, i.e., $\varphi(r)$ for $\bar{R}$ (\ref{eq:conjR})
 is $1+\frac{1}{2}r$ and root of the polynomial has magnitude greater than $1$.
\item One can show that $R$ or $T$ has infinitely many
imaginary-axis zeros if and only if corresponding $\varphi(r)$ has
a root with magnitude $1$ in Lemma~\ref{lemma:finitezeros}. Since
by assumption $A.2$, $P$ has no imaginary axis poles or zeros,
this possibility is eliminated. In fact, if plant $P$ does not
have infinitely many imaginary-axis poles or zeros, the magnitude
of roots of $\varphi(r)$ is never equal to $1$. 
\item 
For a given system $R$, if magnitudes of all roots of $\varphi(r)$
in Lemma \ref{lemma:finitezeros} are smaller than one, then $R$ is
an $I$-system.
\item $R$ is an $F$-system $\Longleftrightarrow$ $\bar{R}$ is an
$I$-system.
\end{enumerate}

\vspace*{-0.5cm}
\subsection{FIR Part of the Time Delay Systems}\label{section:FIRstructure}
We now show a special structure of time delay systems. This key
lemma is used in the next section.
\begin{lemma}
\label{lemma:FIR} \vspace*{-0.25cm} Let $R$ be as in Lemma
\ref{lemma:finitezeros} and $M_R$ be a finite dimensional system
whose zeros are included in the zeros of $R$. Let
$\mathcal{S}_z^+$ be the set of common $\mathbb{C}_+$ zeros of $R$
and $M_R$. Then $\frac{R}{M_R}$, can be decomposed as
$\frac{R}{M_R}=H_R(s)+\mathcal{F}_R(s)$, where $H_R$ is a system
whose poles are outside of $\mathcal{S}_z^+$ and the impulse
response of $\mathcal{F}_R$ has finite support (by a slight abuse
of notation we say $\mathcal{F}_R$ is an FIR filter).
\end{lemma}
\vspace*{-0.25cm} {\em Proof}. For simplicity assume that
$z_1,z_2,\dots,z_{n_z}\in\mathcal{S}_z^+$ are distinct.
We can rewrite $\frac{R}{M_R}$ as $  \frac{R}{M_R}=\sum_{i=1}^n
\frac{R_i}{M_R}e^{-h_i s}$, and decompose each term by partial
fraction, $\frac{R_i}{M_R}=H_i+F_i$ where the poles of $F_i$ are
elements of $\mathcal{S}_z^+$ and define the terms $H_R$ and
$\mathcal{F}_R$ as, \bd H_R(s)=\sum_{i=1}^n H_i(s)e^{-h_i s}, \;
\mathcal{F}_R(s)=\sum_{i=1}^n F_i(s)e^{-h_i s}. \ed where $F_i$ is
strictly proper and $\mathcal{F}_R(z_k)$ is finite $\forall
\;i=1,\dots,n_z$. The lemma ends if we can show that
$\mathcal{F}_R$ is FIR filter. Inverse Laplace transform of
$\mathcal{F}_R$ can be written as,\\
$ f_R(t)=\sum_{k=1}^{n_z}\left[\sum_{i=1}^n Res
\{F_i(s)\}\Big|_{s=z_k}\;e^{z_k(t-h_i)}u_{h_i}(t) \right] $ where
$u_{h_i}(t)=u(t-h_i)$, $u(t)$ and $Res(.)$ are unit step function
and the residue of the function
respectively. For $t>h_n$, we have \\
$ f_R(t)=\sum_{k=1}^{n_z}e^{z_k t}\left[\sum_{i=1}^n
Res\{F_i(s)\}\Big|_{s=z_k}\;e^{-h_i z_k} \right] $. Since,
$Res\{F_i(s)\}\Big|_{s=z_k}=R_i(z_k)Res\{M_R(s)\}\Big|_{s=z_k}$, $
f_R(t)=\sum_{k=1}^{n_z}\left[e^{z_k t}Res\{M_R(s)\}\Big|_{s=z_k}R(z_k)\right]\equiv0$ \\
for $t>h_n$ using the fact $\{z_k\}_{k=1}^{n_z}$ are the zeros of
$R$. Therefore, we can conclude that $\mathcal{F}_R$ is a FIR
filter with support $[0,h_n]$. Note that the above arguments are
also valid for common zeros with multiplicities in
$\mathcal{S}_z^+$. \hfill $\Box$

Note that this decomposition eliminates unstable pole-zero
cancellation in $\frac{R}{M_R}$ and brings it into a form which is
easy for numerical implementation. Lemma~\ref{lemma:FIR} explains
the FIR part of the $\Hi$ controllers as shown below. 
Assume that $R$ is defined as in Definition \ref{def:TDS} and
$R_0$ is a bi-proper, finite dimensional system. By partial
fraction, $ \frac{R_i}{R_0}=R_{i,r}+R_{i,0} \quad  \forall
i=1,\dots,n $, where the $R_{i,0}$ is strictly proper transfer
function whose poles are same as the zeros of $R_0$. Then, the
{\em decomposition operator}, $\Phi$, is defined as, \bd
\Phi(R,R_0)=H_R+\mathcal{F}_R \ed where $H_R=\sum_{i=1}^n
R_{i,r}e^{-h_i s}$ and $\mathcal{F}_R=\sum_{i=1}^n R_{i,0}e^{-h_i
s}$ are infinite dimensional systems. Note that if the zeros of
$R_0$ are also unstable zeros of $R$, then $\mathcal{F}_R$ is a
FIR filter by Lemma \ref{lemma:FIR}.

\vspace*{-0.5cm}
\section{Main Results} \label{section:main}
In this section, we construct the optimal $\Hi$ controller for the
plant $P$, (\ref{eq:seqplant}), satisfying assumptions $A.1-A.4$.
By Corollary \ref{cor:A3A4}, the plant, $P=\frac{R}{T}$, is
assumed to be either IF, FI or FF plant.

For each case, we will find optimal $\Hi$ controller and obtain a
structure where there is no internal unstable pole-zero
cancellation in the controller.

\vspace*{-0.35cm}
\subsection{Factorization of the Plants}
\label{section:plantfactorization} In order to apply the Skew-
Toeplitz approach, we need to factorize the plant as in
(\ref{eq:plantP1}) or (\ref{eq:plantP2}).

 \vspace*{-0.35cm}
\subsubsection{IF Plant Factorization}
Assume that the plant in (\ref{eq:seqplant}) satisfies $A.1-A.4$,
and $R$ is $I$-system and $T$ is $F$-system. Then $P$ is in the
form (\ref{eq:plantP1}), where
\bea \label{eq:IFfactorization}
\nonumber \hat{m}_n&=&e^{-(h_1-\tau_1)s}M_{\bar{R}} \frac{\{e^{h_1
s}R\}}{\bar{R}}, \quad \hat{m}_d=M_T, \\
\hat{N}_o&=&\frac{\bar{R}}{M_{\bar{R}}} \frac{M_T}{\{e^{\tau_1
s}T\}}. \eea where $M_{\bar{R}}$ is an inner function whose zeros
are the unstable zeros of $\bar{R}(s)$. Since $R$ is $I$-system,
conjugate of $R$  has finitely many unstable zeros, so
$M_{\bar{R}}$ is well-defined. Similarly, zeros of $M_T$ are
unstable zeros of $T$. Note that $\hat{m}_n$ and $\hat{m}_d$ are
inner functions, infinite and finite dimensional respectively.
$\hat{N}_o$ is an outer term. 
\subsubsection{FI Plant Factorization}
Let the plant (\ref{eq:seqplant}) satisfy $A.1-A.4$ (with
$h_1=\tau_1=0$), and assume $R$ is $F$-system and $T$ is
$I$-system. Then the plant $P$ can be factorized as in
(\ref{eq:plantP2}),
\bea \nonumber \tilde{m}_n&=&M_{\bar{T}}\frac{T}{\bar{T}}, \quad
\tilde{m}_d=M_R(s), \quad
\tilde{N}_o=\frac{R}{M_R}\frac{M_{\bar{T}}}{\bar{T}}. \eea The
zeros of  $M_R$ are right half plane zeros of $R$. The unstable
zeros of $\bar{T}(s)$ are the same as the zeros of $M_{\bar{T}}$.
Similar to previous section, conjugate of $T$ has finitely many
unstable zeros since $T$ is an $I$-system. The right half plane
pole-zero cancellations in $\tilde{m}_n$ and $\tilde{N}_o$ will be
eliminated in Section~\ref{section:FIcontroller} by the method of
Section~\ref{section:FIRstructure}.
\subsubsection{FF Plant Factorization}
Let $P=R/T$ satisfy $A.1-A.4$, with $R$ and $T$ being $F$-systems.
In this case $P$ is in the from (\ref{eq:plantP1}),
\bea \label{eq:FFfactorization}
\nonumber \hat{m}_n&=&e^{-(h_1-\tau_1)s}M_R, \quad \hat{m}_d=M_T(s), \\
\hat{N}_o&=&\frac{\{e^{h_1s}R\}}{M_R}\frac{M_T}{\{e^{\tau_1s}T\}}
\eea where $M_R$ and $M_T$ are inner functions whose zeros are
unstable zeros of $R$ and $T$ respectively. Note that when
$h_1=\tau_1=0$, $\hat{m}_n$ is finite dimensional. Then, exact
unstable pole-zero cancellations are possible in this case (except
the ones in $\hat{N}_o$).

\vspace*{-0.5cm}
\subsection{Optimal $\Hi$ Controller Design}
\label{section:optHinfdesign}

Optimal $\Hi$ controllers for problems (\ref{eq:wsm1}) and
(\ref{eq:wsm2}) are given in \cite{TO95} and \cite{GO04} for the
plants (\ref{eq:plantP1}) and (\ref{eq:plantP2}) respectively.
Given the plant and the weighting functions, the optimal $\Hi$
cost, $\gamma_{opt}$ can be found as described in these papers.
Then, one needs to compute transfer functions labeled as
$E_{\gamma_{opt}}$, $F_{\gamma_{opt}}$ and $L$. Due to space
limitations we skip this procedure, see \cite{TO95} and
\cite{GO04} for full details. Instead, we now simplify the
structure of the controllers so that a reliable implementation is
possible, i.e. there are no internal unstable pole-zero
cancellations.

\subsubsection{Controller Structure of IF Plants}
\label{section:IFcontroller} By using the method in
\cite{TO95,FOT}, the optimal controller can be written as, \be
\label{eq:CoptIF}
\hat{C}_{opt}=\frac{K_{\hat{\gamma}_{opt}}\left(\frac{e^{\tau_1s}T}{M_T}\right)}
{\frac{\bar{R}}{M_{\bar{R}}}+e^{\tau_1s}RF_{\hat{\gamma}_{opt}}L}
\ee where
$K_{\hat{\gamma}_{opt}}=E_{\hat{\gamma}_{opt}}F_{\hat{\gamma}_{opt}}M_TL$.
In order to obtain this structure of controller: \begin{enumerate}
    \item Do the necessary cancellations in
    $K_{\hat{\gamma}_{opt}}$,
    \item Partition, $K_{\hat{\gamma}_{opt}}$ as,
    $K_{\hat{\gamma}_{opt}}=\theta_{\hat{\gamma}_{opt}}\theta_T$
where $\theta_{\hat{\gamma}_{opt}}$ is a bi-proper transfer
function. The zeros of $\theta_{\hat{\gamma}_{opt}}$ are right
half plane zeros of $E_{\hat{\gamma}_{opt}}M_T$,
    \item By Lemma \ref{lemma:FIR}, obtain ($H_T$, $\mathcal{F}_T$), ($H_{R_1}$,$\mathcal{F}_{R_1}$) and ($H_{R_2}$, $\mathcal{F}_{R_2}$)
    using the partitioning operator,
    \bea
\nonumber H_T+\mathcal{F}_T&=&\Phi(e^{\tau_1s}T\theta_T,M_T), \\
\nonumber
H_{R_1}+\mathcal{F}_{R_1}&=&\Phi(\bar{R},M_{\bar{R}}\theta_{\hat{\gamma}_{opt}}),\\
\nonumber
H_{R_2}+\mathcal{F}_{R_2}&=&\Phi(e^{\tau_1s}RF_{\hat{\gamma}_{opt}}L,\theta_{\hat{\gamma}_{opt}}).
    \eea
\end{enumerate}
Then, the optimal controller has the form, \be
\label{eq:CoptFIRIF}
\hat{C}_{opt}=\frac{H_T+\mathcal{F}_T}{H_{\hat{\gamma}_{opt}}+\mathcal{F}_{\hat{\gamma}_{opt}}}
\ee where $H_T$, $H_{\hat{\gamma}_{opt}}=H_{R_1}+H_{R_2}$ are TDS
and $\mathcal{F}_T$,
$\mathcal{F}_{\hat{\gamma}_{opt}}=\mathcal{F}_{R_1}+\mathcal{F}_{R_2}$
are FIR filters. The controller has no unstable pole-zero
cancellations.
\subsubsection{Controller Structure of FI Plants}
\label{section:FIcontroller} After the data transformation is done
shown as shown in \cite{GO04} $\tilde{\gamma}_{opt}$,
$E_{\tilde{\gamma}_{opt}}$, $F_{\tilde{\gamma}_{opt}}$ and $L$ can
be found as in IF plant case. We can write the inverse of the
optimal controller similar to (\ref{eq:CoptIF}): \be
\label{eq:CoptFI}
\tilde{C}^{-1}_{opt}=\frac{K_{\tilde{\gamma}_{opt}}\left(\frac{R}{M_R}\right)}
{\frac{\bar{T}}{M_{\bar{T}}}+TF_{\tilde{\gamma}_{opt}}L} \ee where
$K_{\tilde{\gamma}_{opt}}=E_{\tilde{\gamma}_{opt}}F_{\tilde{\gamma}_{opt}}M_RL$.
Similar to IF plant case, we can obtain a reliable controller
structure:
\begin{enumerate}
    \item Do the necessary cancellations in
    $K_{\tilde{\gamma}_{opt}}$,
    \item Partition, $K_{\tilde{\gamma}_{opt}}$ as,
    $K_{\tilde{\gamma}_{opt}}=\theta_{\tilde{\gamma}_{opt}}\theta_R$
where $\theta_{\tilde{\gamma}_{opt}}$ is a bi-proper transfer
function. The zeros of $\theta_{\tilde{\gamma}_{opt}}$ are
unstable zeros of $E_{\tilde{\gamma}_{opt}}M_R$,
    \item By Lemma \ref{lemma:FIR}, obtain ($H_R$, $\mathcal{F}_R$), ($H_{T,1}$,$\mathcal{F}_{T_1}$) and
    ($H_{T_2}$, $\mathcal{F}_{T_2}$) using the partitioning operator,
    \bea
\nonumber H_R+\mathcal{F}_R&=&\Phi(R\theta_R,M_R), \\
\nonumber
H_{T_1}+\mathcal{F}_{T_1}&=&\Phi(\bar{T},M_{\bar{T}}\theta_{\tilde{\gamma}_{opt}}),\\
\nonumber
H_{T_2}+\mathcal{F}_{T_2}&=&\Phi(TF_{\tilde{\gamma}_{opt}}L,\theta_{\tilde{\gamma}_{opt}}).
    \eea
\end{enumerate}
Then, the optimal controller has the form, \be
\label{eq:CoptFIRFI}
\tilde{C}_{opt}=\frac{H_{\tilde{\gamma}_{opt}}+\mathcal{F}_{\tilde{\gamma}_{opt}}}{H_R+\mathcal{F}_R}.\ee
where $H_R$, $H_{\tilde{\gamma}_{opt}}=H_{T_1}+H_{T_2}$ are TDS
and $\mathcal{F}_R$,
$\mathcal{F}_{\tilde{\gamma}_{opt}}=\mathcal{F}_{T_1}+\mathcal{F}_{T_2}$
are FIR filters. The controller has no unstable pole-zero
cancellations. Note that the optimal controller is dual case of IF
plants, $R$ and $T$ are interchanged with $h_1=\tau_1=0$.
\subsubsection{Controller Structure of FF Plants} \label{section:FFcontroller}
Structure of FF plants is similar to that of IF plants. We can
calculate $\hat{\gamma}_{opt}$, $E_{\hat{\gamma}_{opt}}$,
$F_{\hat{\gamma}_{opt}}$, $L$ by the method in \cite{TO95,FOT} and
write optimal controller as: \be
\hat{C}_{opt}=\frac{K_{\hat{\gamma}_{opt}}
\frac{\{e^{\tau_1}T\}}{M_T(s)}}{\frac{\{e^{h_1s}R\}}
{M_R}+e^{\tau_1s}RF_{\hat{\gamma}_{opt}}L} \ee where
$K_{\hat{\gamma}_{opt}}=E_{\hat{\gamma}_{opt}}F_{\hat{\gamma}_{opt}}M_TL$.
The optimal $ \Hi$ controller structure can be found by following
similar steps as in IF plants. The controller structure will be
the same as in (\ref{eq:CoptFIRIF}). Note that when
$h_1=\tau_1=0$, since $\hat{m}_n$ in (\ref{eq:FFfactorization}) is
finite dimensional, it possible to cancel the zeros of
$\theta_{\hat{\gamma}_{opt}}$ with denominator.

\vspace*{-0.35cm}
\section{Example}
\begin{figure}[h]
   \centering \includegraphics[width=6cm]{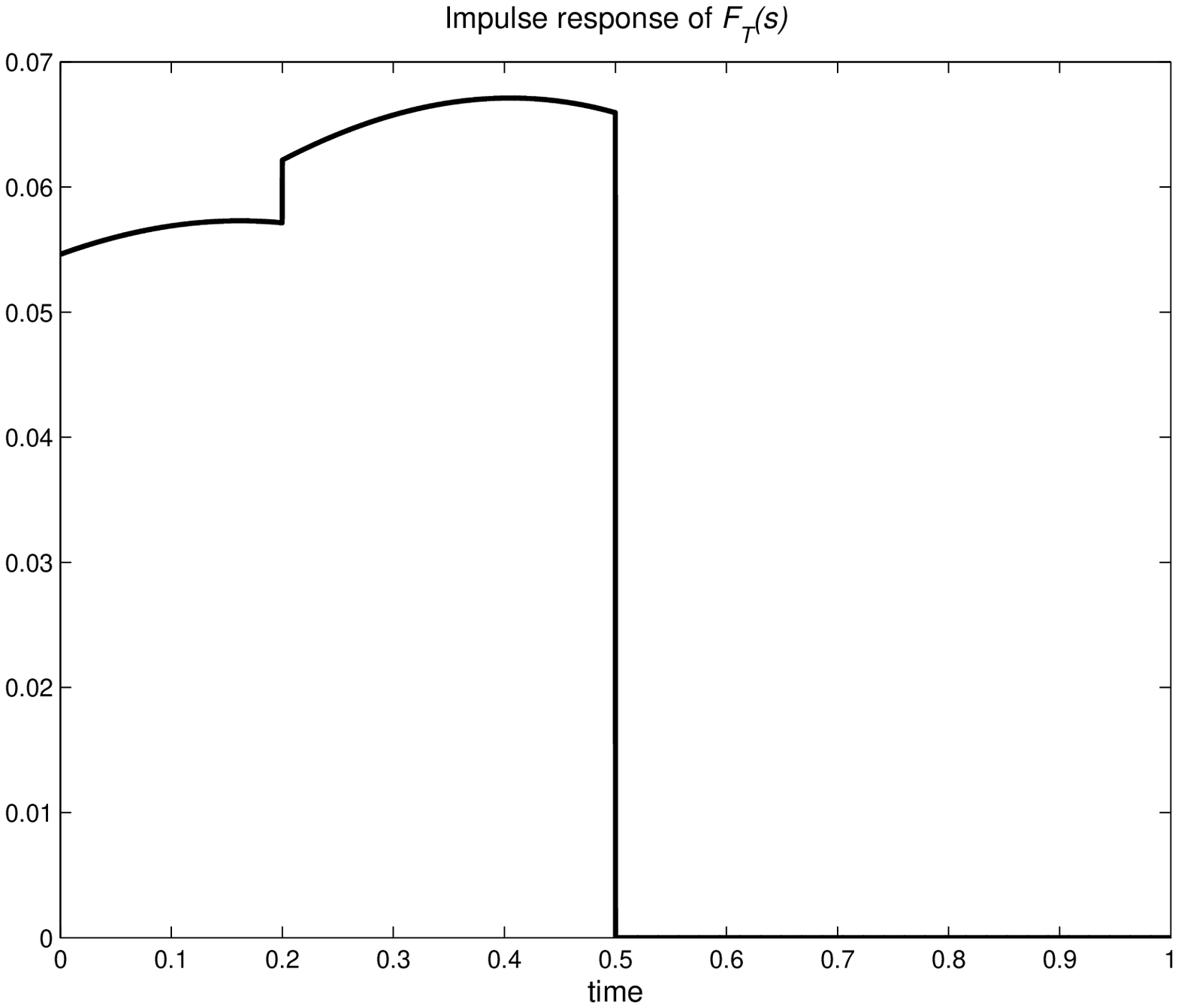}
   \caption{$f_T(t)$} \label{fig:IFFT}
\end{figure}

\begin{figure}[h]
   \centering \includegraphics[width=6cm]{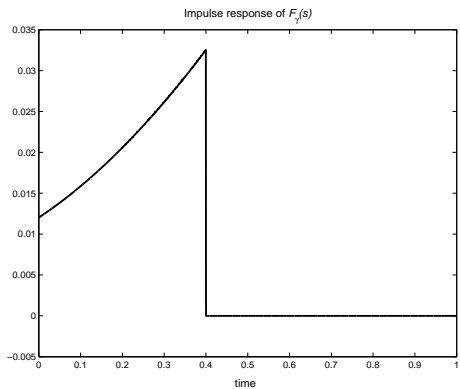}
   \caption{$f_{\hat{\gamma}_{opt}}(t)$} \label{fig:IFFg}
\end{figure}

We consider IF plant (\ref{example:IFplant}) and weights as
$W_1(s)=\frac{2s+2}{10s+1}$ and $W_2(s)=0.2(s+1.1)$. After the
plant is factorized as (\ref{eq:IFfactorization}), the optimal
$\Hi$ cost for two block problem (\ref{eq:wsm1}) is
$\hat{\gamma}_{opt}=0.7203$. The impulse responses of
$\mathcal{F}_T$ and $\mathcal{F}_{\hat{\gamma}_{opt}}$, of the
controller (\ref{eq:CoptFIRIF}), are FIR as in Figures
\ref{fig:IFFT} and \ref{fig:IFFg}, respectively.


\vspace*{-1cm}
\section{Concluding Remarks}
\label{Sec:links}

In this paper we have discussed general time delay systems and
defined FI, IF and FF types of plants. We showed how assumptions
of the Skew Toeplitz theory can be checked, and illustrated
numerically stable implementations of the optimal $\Hi$
controllers, avoiding internal pole-zero cancellations.

We should also mention that if the plant $P$ is written in terms
of specific time delay factors, we may still design optimal $\Hi$
controllers even if $P$ is not in any of the types we have
considered (i.e. IF, FI, and FF types). It is possible to design
$\Hi$ controller for the following cases. Given the plant
$P=\frac{R}{T}$, assume that $T$ is an $F$ system and $R$ is
neither $F$ or $I$ system, but it can be factorized as $R=R_FR_I$
where $R_F$ is an $F$ system and $R_I$ is an $I$ system. Then, we
can factorize the plant as (\ref{eq:plantP1}),
\bea
\nonumber \hat{m}_n&=&M_{R_F}M_{\bar{R}_I}\frac{R_I}{\bar{R}_I}, \quad \hat{m}_d=M_T, \\
\nonumber \hat{N}_o&=&\left(\frac{R_F}{M_{R_F}}\right)
  \left(\frac{\bar{R}_I}{M_{\bar{R}_I}}\right)\left(\frac{M_T}{T}\right)
\eea where $M_{R_F}$, $M_{\bar{R}_I}$ and $M_T$ are finite
dimensional, inner functions whose zeros are unstable zeros of
$R_F$, $\bar{R}_I$ and $T$ respectively. By this factorization,
the optimal controller can be obtained as in~(\ref{eq:CoptFIRIF}).
For the dual case, let $R$ be an $F$ system and $T=T_FT_I$ where
$T_F$ is an $F$ system and $T_I$ is an $I$ system. Now the plant
is in the form (\ref{eq:plantP2}),
\bea
\nonumber \tilde{m}_n&=&M_R,
\quad \tilde{m}_d=M_{T_F}M_{\bar{T}_I}\frac{T_I}{\bar{T}_I}, \\
\nonumber
\tilde{N}_o&=&\left(\frac{R}{M_{R}}\right)
\left(\frac{M_{T_F}}{T_F}\right)\left(\frac{M_{\bar{T}_I}}{\bar{T}_I}\right)
\eea where $M_{R}$, $M_{\bar{T}_I}$ and $M_{T_F}$ are finite
dimensional, inner functions whose zeros are unstable zeros of
$R$, $\bar{T}_I$ and $T_F$ respectively. Now the optimal
controller can be obtained as in~(\ref{eq:CoptFIRFI}).

Another interesting point to note is that the following plant is a
special case of an FF system:
 \bea \label{plant:ssFF}
\nonumber \dot{x}(t)&=&\sum_{i=1}^{n_A}A_ix(t-h_{A,i})
+\sum_{j=1}^{n_b}b_ju(t-h_{b,j}), \\
y(t)&=&\sum_{k=1}^{n_c}c_kx(t-h_{c,k})+du(t-h_d)
\eea
where
$A_i\in\bbr^{n\times n}$, $b_j\in\bbr^{n\times 1}$,
$c_k\in\bbr^{1\times n}$ and $d\in\bbr$. Define
$x(t):=[x_1(t),\ldots,x_n(t)]^T$. The time-delays,
$\{h_{A,i}\}_{i=1}^{n_A}$, $\{h_{b,i}\}_{i=1}^{n_b}$,
$\{h_{c,i}\}_{i=1}^{n_c}$ are nonnegative rational numbers with
ascending ordering respectively and $h_d\geq0$. Therefore, we can
design an optimal $\Hi$ controller for the plant
(\ref{plant:ssFF}) if there are no imaginary axis poles or zeros
(or the weights are chosen in such a way that certain
factorizations in \cite{FOT} can be done).

In general to see the plant type (IF, FI, FF), the transfer
function should be obtained first, then using $R$ and $T$, one can
decide the plant type by Corollary \ref{cor:Fsystem} and
\ref{cor:Isystem}. The optimal $\Hi$ controller can be found by
factorization of the plant and elimination of unstable pole-zero
cancellations.
%

\bibliography{RoConD2006}

\end{document}